%% file: main.tex
\documentclass[11pt,a4paper]{article}
\usepackage{ifthen} % for conditional statements
\newboolean{pdflatex}
\setboolean{pdflatex}{true} % False for eps figures 

\newboolean{articletitles}
\setboolean{articletitles}{true} % False removes titles in references

\newboolean{uprightparticles}
\setboolean{uprightparticles}{false} %True for upright particle symbols

\newboolean{inbibliography}
\setboolean{inbibliography}{false} %True once you enter the bibliography

\usepackage{ifthen}
\setboolean{uprightparticles}{false}
\setboolean{pdflatex}{true}
\usepackage{graphicx}

\usepackage[utf8]{inputenc}
\usepackage{color}
\usepackage{xspace}
\usepackage{siunitx}
\usepackage{amssymb}
\usepackage{amsmath}
\usepackage{amsfonts}
\input{lhcb-symbols-def}

\input{testbeam-symbols-def}

\graphicspath{{./Figures/}}

\input{jinst_preamble}

\input{testbeam-symbols-def}
\usepackage{makecell}

\input{jinst_title}

\begin{document}
\maketitle

%% Uncomment during review phase. 
%% Comment before a final submission.
%\linenumbers

% You can include short sections directly in the main tex file.
% However, for larger papers it is desirable to split the text into
% several semiautonomous files, which can be revised independently.
% This is especially useful when developing a document in
% collaboration with several people, since then different parts can be
% edited independently.  This type of file organization is shown here.
% 
%\input{summary}

%\clearpage 

\section{Introduction} 
\label{sec:introduction} 
\input{01_introduction}

\section{Hardware description} 
\label{sec:hardware} 
\input{02_hardwareDescription}

\section{Data acquisition} 
\label{sec:daq} 
\input{03_daq}

\section{Experiment control system and monitoring} 
\label{sec:slowControlAndMonitoring} 
\input{04_slowControlAndMonitoring}
%\section{Performance} \label{sec:performance} \input{performance}

\section{Spatial resolution and efficiency} 
\label{sec:spacePerformance} 
\input{05_spatialPerformance}

\section{Temporal performance} 
\label{sec:timePerformance} 
\input{06_temporalPerformance}

\section{Conclusions} 
\label{sec:conclusion} 
\input{07_conclusion}

\section{Acknowledgements}
\label{sec:acknowledgements}
\input{08_acknowledgements}

\clearpage
\addcontentsline{toc}{section}{References}
\setboolean{inbibliography}{true}
\bibliographystyle{JHEP}
\bibliography{main}

\end{document}

%% file: lhcb-symbols-def.tex
\usepackage{xspace} 
\usepackage{upgreek}

\def\lhcb   {\mbox{LHCb}\xspace}

\def\MagUp {\mbox{\em Mag\kern -0.05em Up}\xspace}

\ifthenelse{\boolean{uprightparticles}}%
{

 \def\PDelta      {\ensuremath{\Delta}\xspace}                 
 \def\PXi         {\ensuremath{\Xi}\xspace}                 
 \def\PLambda     {\ensuremath{\Lambda}\xspace}                 
 \def\PSigma      {\ensuremath{\Sigma}\xspace}                 
 \def\POmega      {\ensuremath{\Omega}\xspace}                 
 \def\PUpsilon    {\ensuremath{\Upsilon}\xspace}
 \let\oldPi\Pi
 \def\PPi         {\ensuremath{\oldPi}\xspace}

 \def\PB      {\ensuremath{\mathrm{B}}\xspace}                 
                  
 \def\PD      {\ensuremath{\mathrm{D}}\xspace}

 \def\PK      {\ensuremath{\mathrm{K}}\xspace}

 \def\Pe      {\ensuremath{\mathrm{e}}\xspace}

 \def\Pi      {\ensuremath{\mathrm{i}}\xspace}

 \def\Ps      {\ensuremath{\mathrm{s}}\xspace}

 \def\thebaroffset{0.0em}
}
{

 \mathchardef\PDelta="7101
 \mathchardef\PXi="7104
 \mathchardef\PLambda="7103
 \mathchardef\PSigma="7106
 \mathchardef\POmega="710A
 \mathchardef\PUpsilon="7107
 \mathchardef\PPi="7105
                  
 \def\PB      {\ensuremath{B}\xspace}                 
                  
 \def\PD      {\ensuremath{D}\xspace}

 \def\PK      {\ensuremath{K}\xspace}

 \def\Pe      {\ensuremath{e}\xspace}

 \def\Pi      {\ensuremath{i}\xspace}

 \def\Ps      {\ensuremath{s}\xspace}

 \def\thebaroffset{0.18em}
}
\newcommand{\offsetoverline}[2][\thebaroffset]{\kern #1\overline{\kern -#1 #2}}%

\makeatletter
\ifcase \@ptsize \relax% 10pt
  \newcommand{\miniscule}{\@setfontsize\miniscule{4}{5}}% \tiny: 5/6
\or% 11pt
  \newcommand{\miniscule}{\@setfontsize\miniscule{5}{6}}% \tiny: 6/7
\or% 12pt
  \newcommand{\miniscule}{\@setfontsize\miniscule{5}{6}}% \tiny: 6/7
\fi
\makeatother

\DeclareRobustCommand{\optbar}[1]{\shortstack{{\miniscule (\rule[.5ex]{1.25em}{.18mm})}
  \\ [-.7ex] $#1$}}

\def\en         {{\ensuremath{\Pe^-}}\xspace}   % electron negative (\em is taken)

\def\squark    {{\ensuremath{\Ps}}\xspace}

\def\KorKbar {\kern \thebaroffset\optbar{\kern -\thebaroffset \PK}{}\xspace}

\def\D       {{\ensuremath{\PD}}\xspace}

\def\DorDbar {\kern \thebaroffset\optbar{\kern -\thebaroffset \PD}\xspace}

\def\Dp      {{\ensuremath{\D^+}}\xspace}
\def\Dm      {{\ensuremath{\D^-}}\xspace}

\def\DpDm    {\ensuremath{\Dp {\kern -0.16em \Dm}}\xspace}

\def\B       {{\ensuremath{\PB}}\xspace}

\def\BorBbar {\kern \thebaroffset\optbar{\kern -\thebaroffset \PB}\xspace}

\def\Bd      {{\ensuremath{\B^0}}\xspace}

\def\BdorBdbar {\kern \thebaroffset\optbar{\kern -\thebaroffset \Bd}\xspace}

\def\Bs      {{\ensuremath{\B^0_\squark}}\xspace}

\def\BsorBsbar {\kern \thebaroffset\optbar{\kern -\thebaroffset \Bs}\xspace}

\def\Y#1S{\ensuremath{\PUpsilon{(#1S)}}\xspace}

\def\LorLbar     {\kern \thebaroffset\optbar{\kern -\thebaroffset \PLambda}\xspace}

\def\AT#1     {\ensuremath{A_{\mathrm{T}}^{#1}}\xspace}           % 2

\def\C#1      {\ensuremath{\mathcal{C}_{#1}}\xspace}                       % 9
\def\Cp#1     {\ensuremath{\mathcal{C}_{#1}^{'}}\xspace}                    % 7
\def\Ceff#1   {\ensuremath{\mathcal{C}_{#1}^{\mathrm{(eff)}}}\xspace}        % 9  
\def\Cpeff#1  {\ensuremath{\mathcal{C}_{#1}^{'\mathrm{(eff)}}}\xspace}       % 7
\def\Ope#1    {\ensuremath{\mathcal{O}_{#1}}\xspace}                       % 2
\def\Opep#1   {\ensuremath{\mathcal{O}_{#1}^{'}}\xspace}                    % 7

\newcommand{\nospaceunit}[1]{\ensuremath{\text{#1}}}       
\newcommand{\aunit}[1]{\ensuremath{\text{\,#1}}}       
\newcommand{\tev}{\aunit{Te\kern -0.1em V}\xspace}
\newcommand{\gev}{\aunit{Ge\kern -0.1em V}\xspace}
\newcommand{\mev}{\aunit{Me\kern -0.1em V}\xspace}
\newcommand{\kev}{\aunit{ke\kern -0.1em V}\xspace}
\newcommand{\ev}{\aunit{e\kern -0.1em V}\xspace}
 
\newcommand{\mevc}{\ensuremath{\aunit{Me\kern -0.1em V\!/}c}\xspace}
\newcommand{\gevc}{\ensuremath{\aunit{Ge\kern -0.1em V\!/}c}\xspace}
\newcommand{\mevcc}{\ensuremath{\aunit{Me\kern -0.1em V\!/}c^2}\xspace}
\newcommand{\gevcc}{\ensuremath{\aunit{Ge\kern -0.1em V\!/}c^2}\xspace}
\def\cm   {\aunit{cm}\xspace}

\def\mm   {\aunit{mm}\xspace}
\def\mma  {\ensuremath{\aunit{mm}^2}\xspace}
\def\mum  {\ensuremath{\,\upmu\nospaceunit{m}}\xspace}
\def\muma {\ensuremath{\,\upmu\nospaceunit{m}^2}\xspace}

\def\ns   {\ensuremath{\aunit{ns}}\xspace}
\def\ps   {\ensuremath{\aunit{ps}}\xspace}

\def\mhz  {\ensuremath{\aunit{MHz}}\xspace}

\def\gsim{{~\raise.15em\hbox{$>$}\kern-.85em
          \lower.35em\hbox{$\sim$}~}\xspace}
\def\lsim{{~\raise.15em\hbox{$<$}\kern-.85em
          \lower.35em\hbox{$\sim$}~}\xspace}

\def\degrees{\ensuremath{^{\circ}}\xspace}

\def\gbps        {\aunit{Gbit/s}\xspace}

\def\tell1  {TELL1\xspace}
\def\ukl1   {UKL1\xspace}

%% file: testbeam-symbols-def.tex
\def\ttt {time-to-threshold\xspace}

\def\gbps{\ensuremath{\aunit{Gbps}}\xspace}

\def\kepler {\textsc{Kepler}\xspace}

%% file: jinst_preamble.tex
\usepackage{microtype}
\usepackage{xspace} % To avoid problems with missing or double spaces after
\usepackage{caption} %these three command get the figure and table captions automatically small

\usepackage{graphicx}  % to include figures (can also use other packages)
\usepackage{color}
\usepackage{colortbl}
\graphicspath{{./figs/}} % Make Latex search fig subdir for figures
\usepackage{pdflscape}

\usepackage{siunitx} %proper units
\usepackage{amsmath} % Adds a large collection of math symbols
\usepackage{amssymb}
\usepackage{amsfonts}
\usepackage{upgreek} % Adds in support for greek letters in roman typeset

\usepackage{jinstpub}

\usepackage{natbib}
\usepackage{graphicx}
\usepackage{siunitx}
\usepackage{physics}
\usepackage{cleveref}
\usepackage{tikz}
\usepackage{tikz-3dplot}
\usetikzlibrary{arrows.meta}
\usepackage[nobiblatex]{xurl}

%% file: jinst_title.tex
\title{Reconstruction of charged tracks with Timepix4 ASICs}

\author[a,1]{K.~Akiba\note{Corresponding author},}
\author[a]{M.~van~Beuzekom,}
\author[a]{V.~van~Beveren,}
\author[b,c]{W.~Byczynski,}
\author[b]{V.~Coco,}
\author[b]{P.~Collins,}
\author[d]{E.~Dall'Occo,}
\author[b]{R.~Dumps,}
\author[e]{T.~Evans,}
\author[a]{R.~Geertsema,}
\author[b]{E.~L.~Gkougkousis,}
\author[b]{M.~M.~Halvorsen,}
\author[a]{B.~van~der~Heijden,}
\author[a]{K.~Heijhoff,}
\author[f]{E.~Lemos~Cid,}
\author[g]{T.~Pajero,}
\author[b,d]{D.~Rolf,}
\author[b]{H.~Schindler}

\affiliation[a]{{Nikhef, Science Park 105, 1098 XG Amsterdam, the Netherlands}}
\affiliation[b]{{CERN, 1211 Geneve, Switzerland}}
\affiliation[c]{{Tadeusz Kosciuszko Cracow University of Technology, Cracow, Poland}}
\affiliation[d]{{TU Dortmund, Otto-Hahn-Strasse 4, 44227 Dortmund, Germany}}
\affiliation[e]{{Department of Physics and Astronomy, University of Manchester, Manchester, United Kingdom}}
\affiliation[f]{{Instituto Galego de Fisica de Altas Enerxias (IGFAE), Universidade de Santiago de Compostela, Santiago de Compostela, Spain}}
\affiliation[g]{{Department of Physics, University of Oxford, Denys Wilkinson Bldg., Keble Road, Oxford, OX1 3RH, United Kingdom}}

\emailAdd{kazu.akiba@nikhef.nl}

\abstract{

The spatial and temporal performance of a four-plane system composed of silicon sensors bump-bonded to Timepix4 ASICs is assessed with a $180\gevc$ mixed hadron beam at the CERN SPS and reported in detail.
Particle tracks are reconstructed using time-space measurements from the four detector planes, two 100\mum planes perpendicular to the beam and two 300\mum sensors under an angle of 9\degrees.
The spatial hit resolution is assessed to be  $(15.5\pm 0.5)\mum$ and $(4.5\pm0.3)\mum$ for $100\mum$ and $300\mum$ thick sensors, respectively.  
The timestamps from the detectors are also measured with fine precision, yielding time resolutions of $(452\pm10)\ps$, $(420\pm10)\ps$, $(639\pm10)\ps$, $(631\pm10)\ps$ for the two 100\mum and two 300\mum thick sensors respectively. 
These measurements are combined to a track time resolution of $(340\pm 5)\ps$.
The design of the detector system is described together with its data acquisition system, operational infrastructure, and dedicated software.

}

\keywords{Solid state detectors; Particle tracking detectors (Solid-state detectors); Hybrid detectors; Timing detectors}

%% file: 01_introduction.tex
Future experiments in high energy physics will require timing measurements of the order of 10~ps  in addition to the state-of-the-art spatial measurements. 
The main motivation is to cope with the high occupancy at hadron colliders operating at a high number of collisions  per bunch crossing, by separating tracks from different quasi-simultaneous collisions~\cite{LHCbVELOgroup:2022}.
The Timepix Application Specific Integrated Circuit (ASIC) family has previously been employed in the reconstruction of charged particle trajectories~\cite{Akiba:2019faz,Heijhoff:2020mlk,Heijhoff:2021rtu,DallOcco:2021tjb}, in particular as an R\&D platform for sensors, ASICs and other detector components used for the upgrades of the \lhcb experiment.  
Timepix4~\cite{tpx4_jinst} is a novel ASIC designed for performing both temporal and spatial measurements with $195\ps$ bin width  and $55 \times 55 \muma$ pixel size.
Its timing precision enables the use of spatial and temporal information in a 4D-tracking approach, 
which is essential in the 
R\&D efforts for the next generation of experiments.

In this paper the design of a single arm four-plane telescope based on the Timepix4v1 ASIC is described together with the data acquisition system, operational infrastructure and dedicated software.
This is a first step towards a two arm telescope with at least eight planes with the final version of Timepix4 ASIC~\cite{llopart2022timepix4}, targeting
a spatial resolution of 2\mum or better and a temporal resolution of O(30)~ps.
Finally, the  spatial and temporal performances are assessed using a $180\gevc$ mixed hadron beam  at the SPS H8 beam line facility~\cite{sps-h8}.

%% file: 02_hardwareDescription.tex
\begin{figure}[tb]
	\centering
	\includegraphics[width=0.9\textwidth]{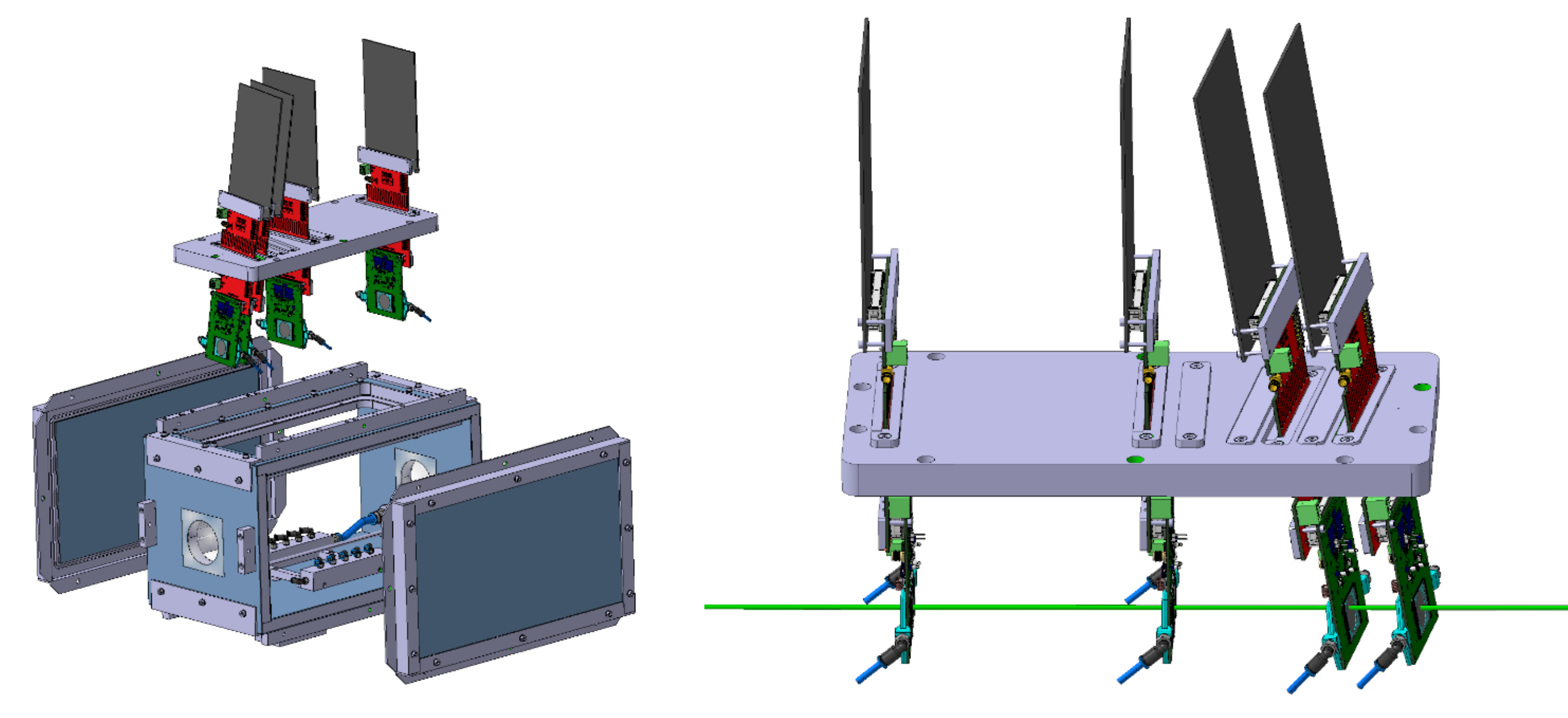}
	\caption{Mechanical design of the telescope-like arrangement of four measuring planes.
	The sensors are placed in a light-tight nitrogen environment, separated from the outside by flanges placed at predefined positions.
	The green solid cylinder line cutting through the planes represents the traversing beam.}
	\label{fig:telescope}
\end{figure}

The telescope consists of a single arm with four detector planes as illustrated in \cref{fig:telescope}. 
A global right-handed coordinate frame is defined with the $z$ axis in the direction of the beam and the $y$ axis pointing upwards. 
This convention is adopted throughout this paper. 

The detectors are mounted inside a custom hermetic enclosure to provide a cold, light-tight and humidity free environment.
The top cover of this box was machined with slots to allow the insertion of detector planes with the use of matching flanges. 
The individual flanges are composed of matching half-moons which are attached to the detector boards for insertion in the slots.
The positions of the telescope planes along the $z$ axis are determined by predefined slots on the top cover, and are 0, 150, 250 and $290\mm$.
The slots are machined to achieve   different angles of the sensor planes with respect to the $z$ axis. 
The two upstream sensors are perpendicular to the $z$ axis to achieve a better temporal resolution. 
The other two sensors are angled at 9\degrees with respect to the $x$ and $y$ axes in order to improve the spatial resolution~\cite{emma}.
For the majority of the data collection period, the first two slots were instrumented with $100\mum$ thick sensors (with identifiers N30 and N29), while $300 \mum$ thick sensors (identified by N23 and N28) occupied the downstream slots.
In the following sections, this is referred to as the default configuration.
A limited data set was also acquired with an alternative configuration, where one $100\mum$ sensor (N29) was placed in an angled slot and a $300 \mum$ sensor (N23) in the perpendicular slot.
The base of the telescope box is mounted on a remote controlled motion stage, which allows the entire telescope to be moved along the $x$ and $y$ axes, to  align the telescope with respect to the beam.

\subsection{Timepix4 ASIC}
Timepix4 is a readout ASIC capable of simultaneous time-of-arrival (ToA) and time-over-threshold (ToT) measurements~\cite{tpx4_jinst}.
The ASIC has a pixel matrix of $448 \times 512$ square pixels of $55 \mum$ pitch. Hence, the total active area of the detector assemblies is around $24.6 \times 28.2 \mma$.
The ToA of each particle hit above a predefined and programmable threshold is measured by a time-to-digital converter (TDC) with a nominal bin width of $195\ps$. 
Each group of $2\times 4$ pixels, referred to as a superpixel, shares a Voltage Controlled Oscillator (VCO), which provides the $640\mhz$ reference clock for the pixel TDCs. Four versions of this clock are generated, and equally shifted in phase. By registering the state of all four clocks, the TDC bin of 195\ps is achieved.
For this beam test, the first submission of the Timepix4 was used, 
which presents a flaw in the design of the VCO, 
causing it to oscillate about 25\% too fast. 
For the same reason, the control voltage that is generated by the periphery Phase-Locked Loops (PLLs) could not be used, and hence the oscillation frequency was not stabilised, which  degrades the time resolution.
%\footnote{This design flaw is fixed in version 2 of the chip.}.
The ToT measurements used in the analyses presented in this paper are performed with a $25\ns$ bin width.

The Timepix4 ASIC is divided into two halves, denoted top and bottom,  in order to increase readout speeds by placing serialisers on both sides.
The data can be read out by up to 16 serialisers capable of running at a maximum  bandwidth of $10\gbps$ each, to be capable of reading out a maximum hit rate of \SI{3.6}{Mhits\per\milli\meter\squared\per\second}. 
During the beam test, only one serialiser per side was used, and the combined link speed was set to $2\times 2.56\gbps$, thereby limiting the bandwidth to order 100~Mhits/s, which is still about two orders of magnitude larger than the typical rate required for the H8 beam line.

\subsection{Sensors}
\begin{figure}[tb]
	\centering
	\includegraphics[width=0.48\textwidth
%	 ,natheight=400, natwidth =400
	 ]{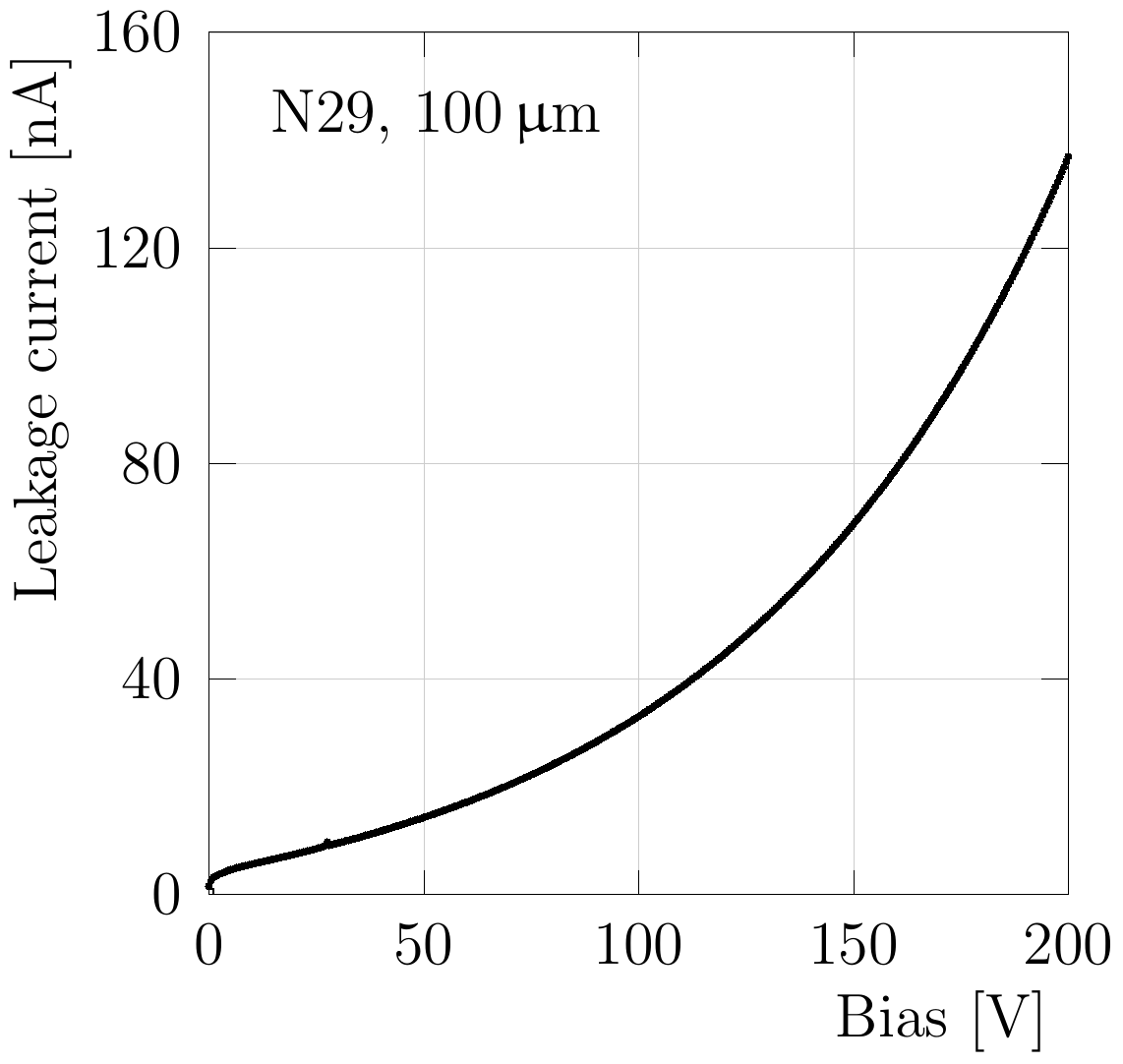}
	\includegraphics[width=0.48\textwidth
%	 ,natheight=400, natwidth =400
	]{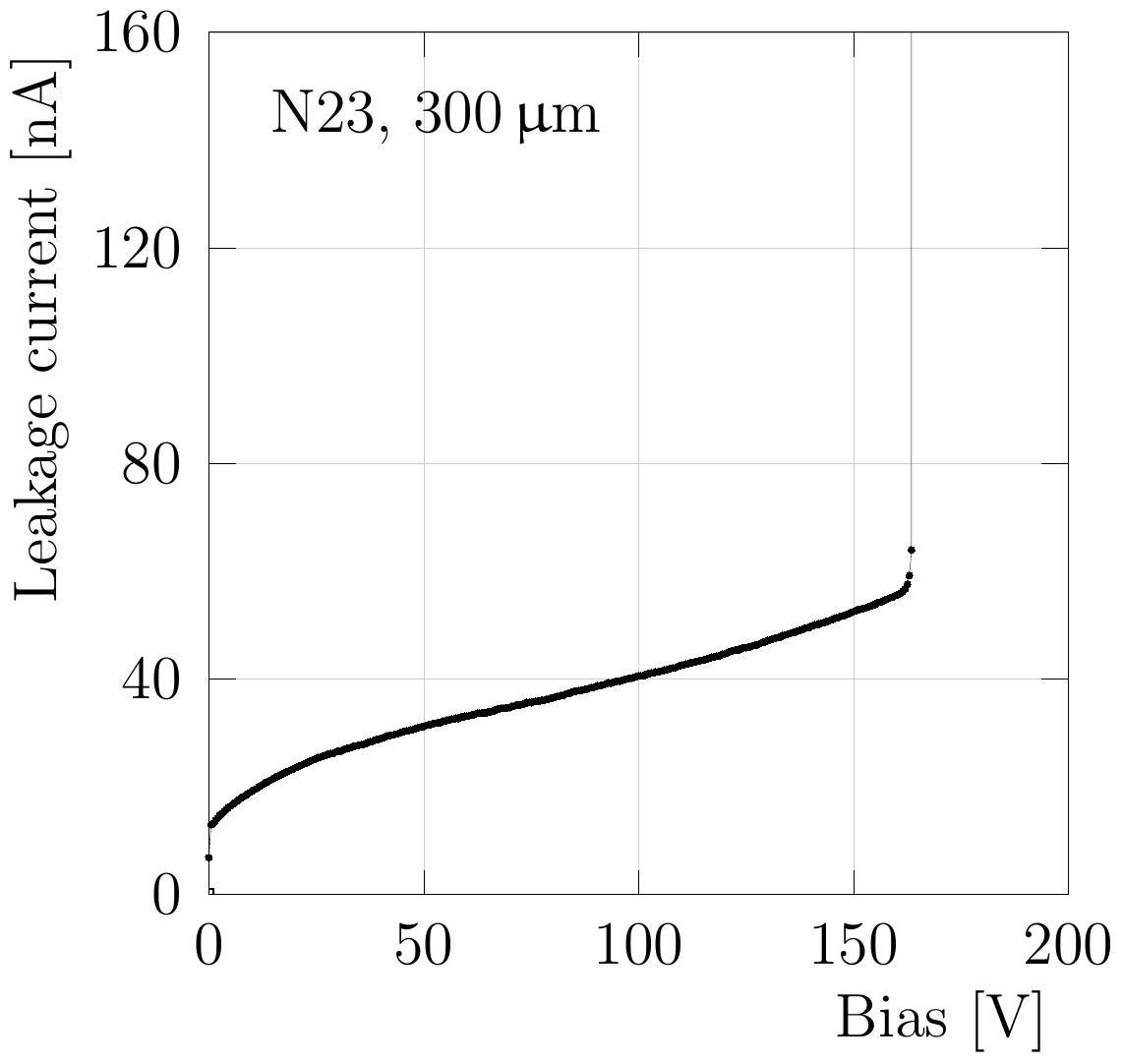}
	\caption{Left (Right): I-V characteristic curve for N29 (100 $\mum$) and N23 (300$\mum$) sensors. The sensors N23, N28, N30 breakdown at 160 V, 140 V and 50 V respectively, while N29 did not breakdown up to 200 V, which was the limit of the measuring range. }
	\label{fig:ivs}
\end{figure}

Planar n-on-p (electron collecting) silicon sensor technology is used in this system.
The sensors are composed of $p$-type silicon bulk with $n^+$-type implants, and were manufactured by ADVACAM.\footnote{Advacam, Tietotie 3, 02150 Espoo, Finland.} 
The back side is a uniform $p^+$ implant which is subsequently metallised to allow for the application of a reverse bias voltage to the sensor.
The front side is segmented with 448 $\times$ 512  approximately $39\mum$ square $n^+$ implants, separated by a uniform p-spray, and covered with under-bump metallisation which allows the pixels sensors to be bonded with solder bumps to the ASICs.
The $300\mum$ sensors are fully depleted at a reverse bias voltage of approximately $50$~{V} with a leakage current of around $15$~{nA} at room temperature, and they could be operated up to $150$~{V} without breakdown. 
The $100\mum$ thick sensors are fully depleted at around $10$~{V} with a leakage current of about $5$~{nA} at room temperature. 
One of the two thin sensors presents breakdown below $50$~{V}, while the other could be reliably biased up to about $200$~{V}. 
Two I-V characteristic curves of the $300\mum$ and $100\mum$ thick sensors are show in \cref{fig:ivs}.

\subsection{Cooling}  
Cooling of the planes is provided by a cooling  block directly attached to the detector board, with a small surface overlap with the ASICs. 
The cooling blocks are made of 3D printed titanium with hollow cavities which allow liquid glycol to circulate through.
The fluid is distributed in parallel to each of the planes.
The cooling blocks have a circular cut-out to minimise the amount of material traversed by incident particles. 
The interface between the detector board and its cooling block was improved by attaching a high thermal conductivity sheet. 
The cooling fluid is pumped through the cooling block by an off-the-shelf chiller.

\subsection{Scintillators} 
The timing measurements are complemented by three plastic (EJ200  \footnote{ELJEN technology, 1300 W. Broadway, Sweetwater, TX 79556, USA. 
%\url{https://eljentechnology.com/products/plastic-scintillators/ej-200-ej-204-ej-208-ej-212}
}) scintillators mounted onto the telescope box. 
Two are placed upstream of the pixel sensors and spaced approximately $2\cm$ apart from each other, while the third is placed at the downstream side. 
The scintillators are instrumented with HPK\footnote{Hamamatsu Photonics K.K., 325-6, Sunayama-cho, Naka-ku, Hamamatsu City,
Shizuoka Pref., 430-8587, Japan.} Photo Multiplier Tubes (PMTs) and their signals are processed by ORTEC-584 constant fraction discriminators (CFD) to minimise the contribution of timewalk to the electronics jitter.
Each CFD output is fed back to a different Timepix4 plane where it is timestamped with a TDC of the same precision as that of the pixels. 
The synchronisation between the ASICs was found to be insufficiently stable to combine the three timestamps.
The individual scintillators are all determined to have a resolution of around 100\ps, therefore the one most upstream was arbitrarily chosen to provide the reference time measurement.

%% file: 03_daq.tex
The Timepix4 ASICs are configured and read out with a custom developed system called SPIDR4, 
which is based on a Xilinx Zynq 7000 FPGA, provides the slow control interface to the Timepix4 via the on-chip ARM processor, which receives configuration commands via a 1 Gbit copper ethernet link.
Regarding the slow control, all SPIDR4 systems are connected to the same computer, which runs four instances of the slow control application, one for each SPIDR4 plus Timepix4. 
Each instance of the DAQ (Data Acquisition) application is controlled by its corresponding slow control application. 
The main DAQ interface to the telescope is managed through a run-control application, which also directs all of the slow control instances.

The pixel data from Timepix4 consists of a 64 bit word for each hit. 
This hit data is transmitted from the chip to the FPGA using a serial 64/66 standard encoding scheme to allow for clock recovery and transmission line balancing. The distance between Timepix4 chip and FPGA is about $25\cm$; the distance could be increased to about one meter, via commercially available FMC cables. 
The Timepix4 is operated with only one 2.56 \gbps serial link per half of the chip, as the track rates at this test beam were relatively low, typically below a million per second.
The data from both links of each Timepix4 device are descrambled by the FPGA in SPIDR4 and packed into UDP datagrams, which are transmitted via an optical 10 Gbit ethernet connection to the DAQ computers, one for each SPIDR4.
The main task of the DAQ application is to write the data to local disk, and no significant data processing is performed.

\subsection{Software}

A software application based on the \textsc{Gaudi} event processing framework~\cite{Clemencic_2010}, \kepler, has been developed for the reconstruction and analysis of data recorded with Timepix telescopes~\cite{Akiba:2019faz}.
The core functionality of the software, which is to provide reconstructed and aligned tracks in a variety of formats to end users, remains largely unchanged.
The main new feature in \kepler is the implementation of a decoder for the Timepix4 data format.
In addition, large improvements to the CPU performance of the reconstruction have been achieved by simplifying the intermediate data structures used by the software and modernisation of the code base. 

\subsection{Data quality monitoring}
\label{sect:monitoring}

A new graphical user interface is implemented to control the execution of \kepler and to monitor the quality of the collected data in real time,  implemented using the \textsc{Qt5} toolkit.
The communication between the interface and the  \kepler server is established through the Distributed Information Management (DIM) protocol~\cite{Gaspar:2001fbw}.
The monitored information mostly consists of histograms of quantities such as the spatial and ToT distributions of the hits in each plane, as well as properties related to the clusters or tracks.
In addition the number of errors in the configuration of the ASICs and in the data communication are displayed.

%% file: 04_slowControlAndMonitoring.tex
\begin{figure}[tb]
	\centering
	\includegraphics[width=\textwidth
%	 ,natheight=400, natwidth =400
	]{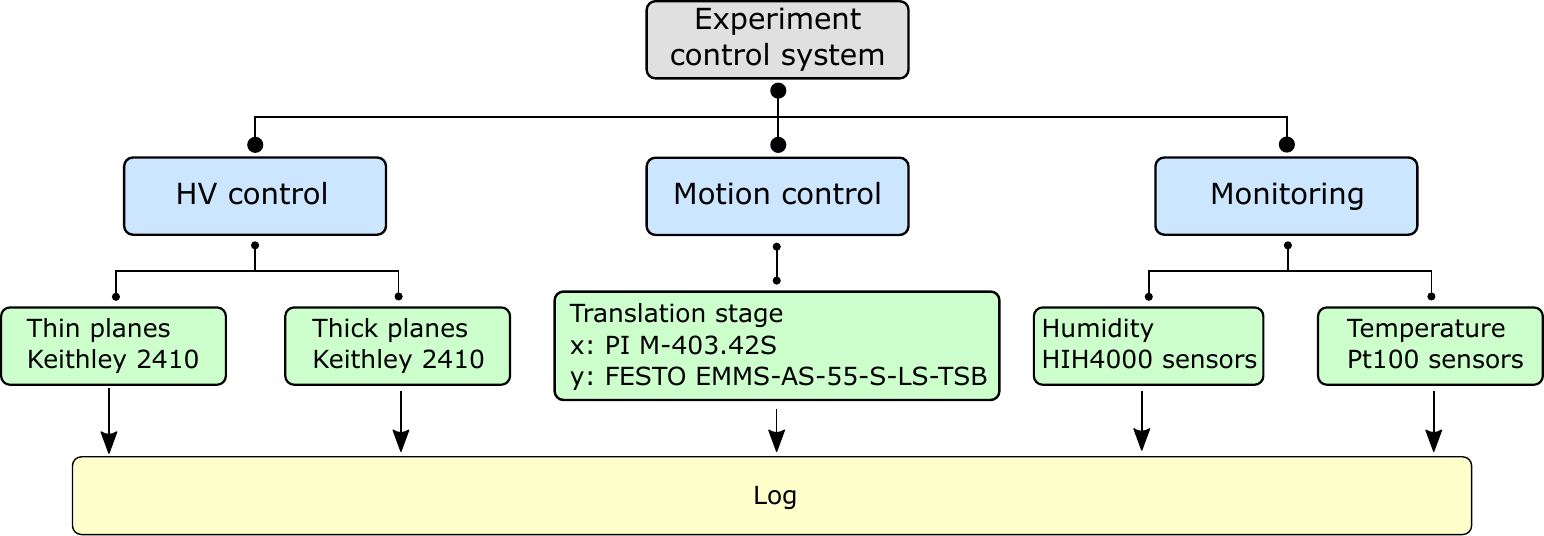}
	\caption{Schematic overview of the experiment control system. }
	\label{fig:slow_control}
\end{figure}

A dedicated experiment control system is implemented to remotely operate motion stages and power supplies, as well as to monitor the environmental conditions of the telescope.
The system implementation is divided in the following way: 
the operation of High Voltage and the monitoring of bias currents (HV control); the operation of the motion stage (motion control); the monitoring of temperature and humidity.  
A block diagram representation of the system is given in~\cref{fig:slow_control}.
The WinCC Open Architecture (OA) software suite~(WinCC OA\footnote{SIMATIC WinCC Open Architecture Portal \url{https://www.winccoa.com/}}) is used to implement the control system, which also provides alarm and logging capabilities.
The communication between WinCC OA and the hardware is established with a custom server based on the DIM protocol and the Open Platform Communications Unified Architecture~(OPC~UA).

The HV-control operates two Keithley 2410 Source Meters~\footnote{Tektronix, Inc. 
14150 SW Karl Braun Drive, Beaverton, OR 97077.  United States} that provide independent bias voltages to the $100\mum$ and $300\mum$ thick sensor planes in the default configuration.  
The translation of the telescope along the $x$ and $y$ axes is performed by PI~\footnote{Physik Instrumente (PI) GmbH \& Co. KG Auf der Roemerstrasse 1 76228 , Karlsruhe, Germany} motion stages 
with a repeatability of $2\mum$.

The temperatures of each plane, as well as the temperature and humidity within the telescope enclosure, are monitored with four-wire Pt100 and HIH4000 sensors~\footnote{Honeywell, Charlotte, North Carolina.}, connected via an Embedded Local Monitor Board (ELMB).
The monitored values for each component are logged in order to enable studies of the telescope performance as a function of  environmental conditions.
In addition, the logging of operational settings such as the bias voltage complements the information manually recorded in the logbook of the testbeam.

%% file: 05_spatialPerformance.tex
Clusters are reconstructed by grouping nearby hits that are within $100\ns$ from each other. This window was selected to ensure that no two tracks overlap in time with the beam rate at SPS, and ensures that all hits with a large timewalk can also be assigned to the correct cluster.
The timestamp of the cluster is assigned as the earliest time measurement in the group of hits~\cite{Heijhoff:2020mlk}.
The cluster position is calculated as the ToT-weighted average of the position of the hits constituting the cluster. 
Particle tracks are reconstructed requiring a cluster in each plane and their trajectories determined using a straight line fit.
The clusters are required to be within a $100\ns$ time interval allowing a low background and high-efficiency track reconstruction given the particle rate at the SPS was less than \SI{2e5}{particles\per\second}. 
The positions and orientations of the planes with respect to each other are determined using the Millepede algorithm~\cite{Blobel:2006yh}, using a set of around 12,000 tracks.
The alignment procedure is repeated several times, with progressively more stringent requirements on the $\chi^2$ of the tracks in each iteration.

\begin{figure}
  \centering
  \includegraphics[width=0.48\textwidth
%	 ,natheight=400, natwidth =400
  ]{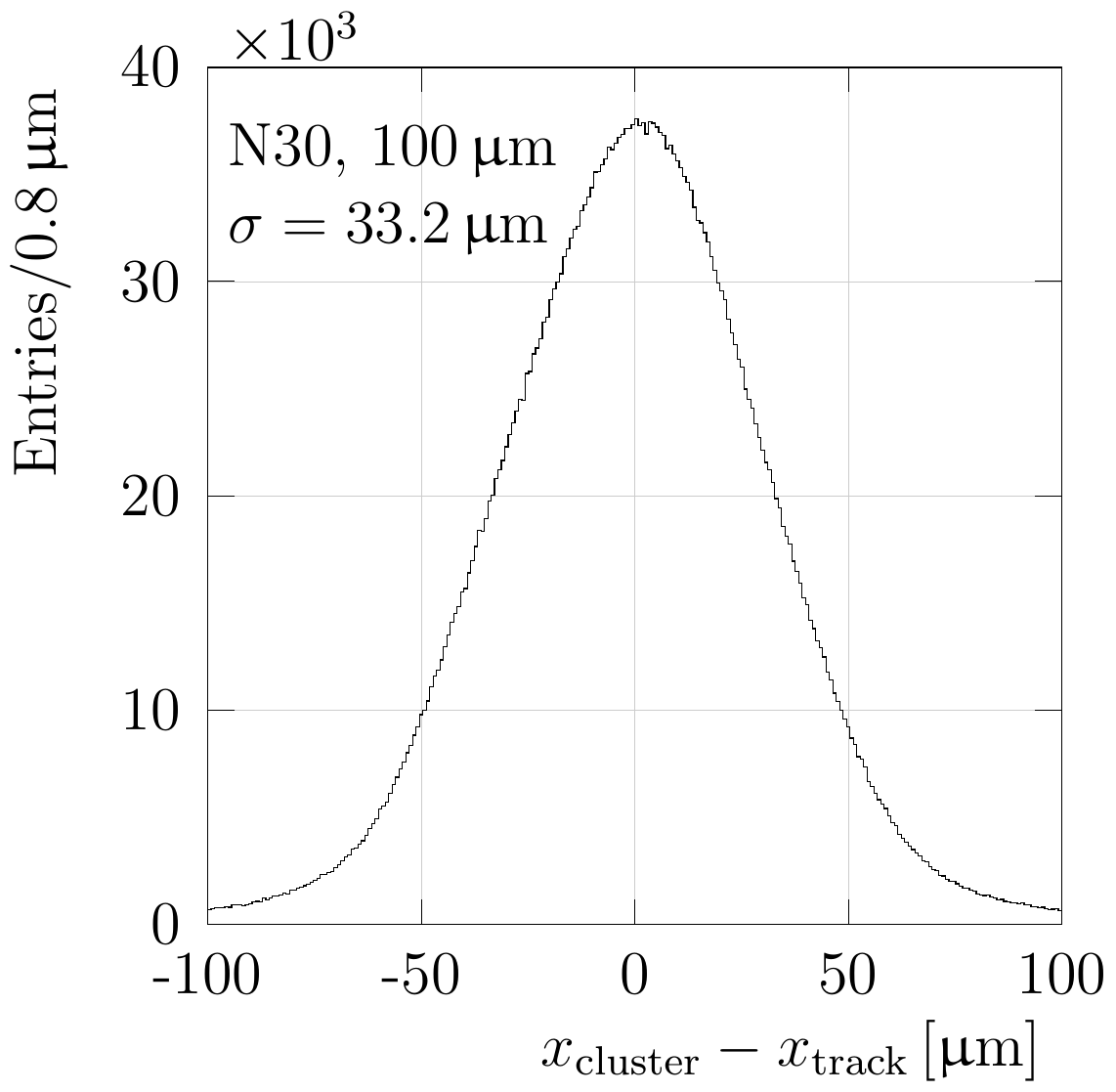}
  \includegraphics[width=0.48\textwidth
%	 ,natheight=400, natwidth =400
  ]{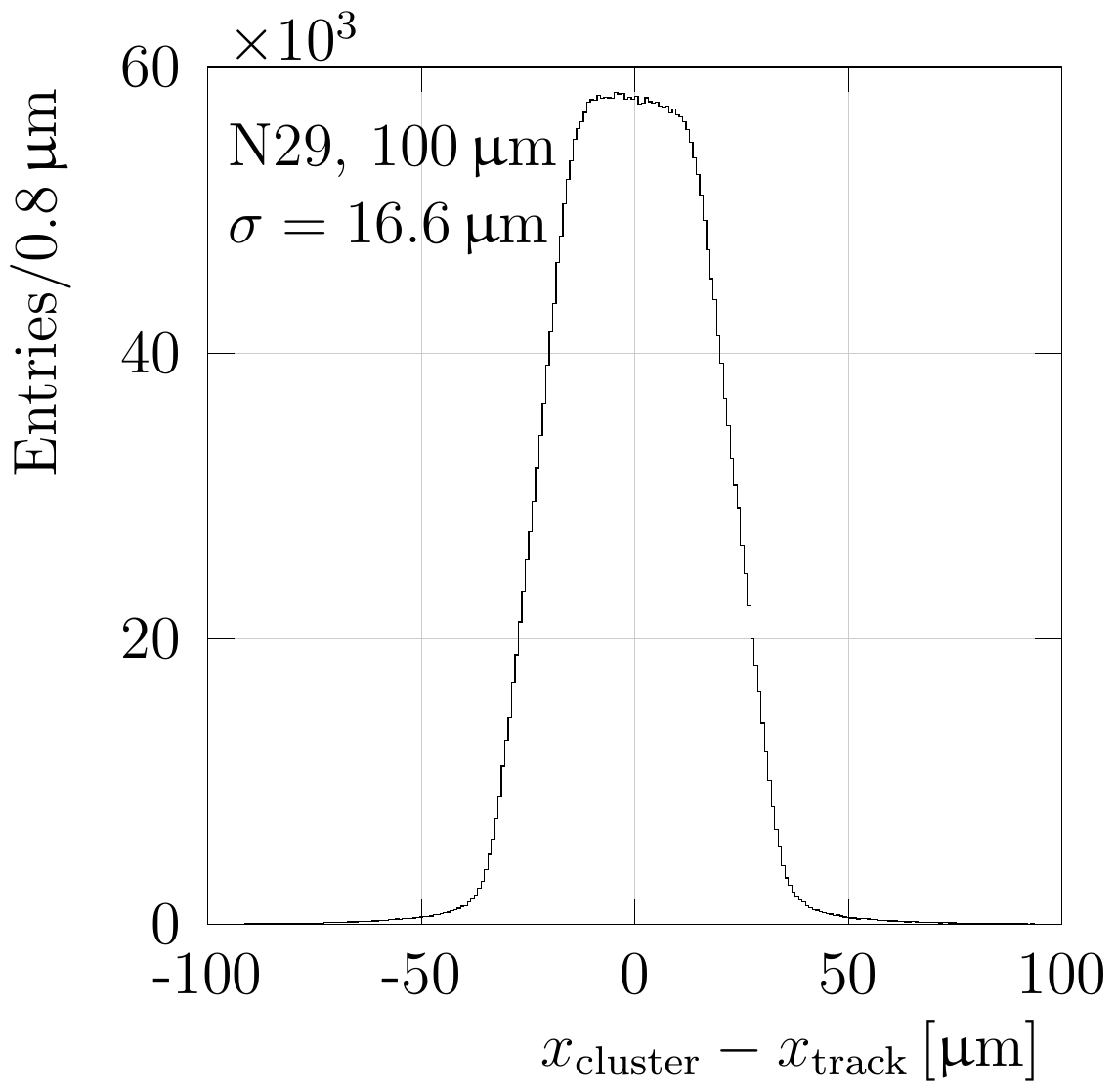}
  \vspace{2mm}
  \includegraphics[width=0.48\textwidth
%	 ,natheight=400, natwidth =400
  ]{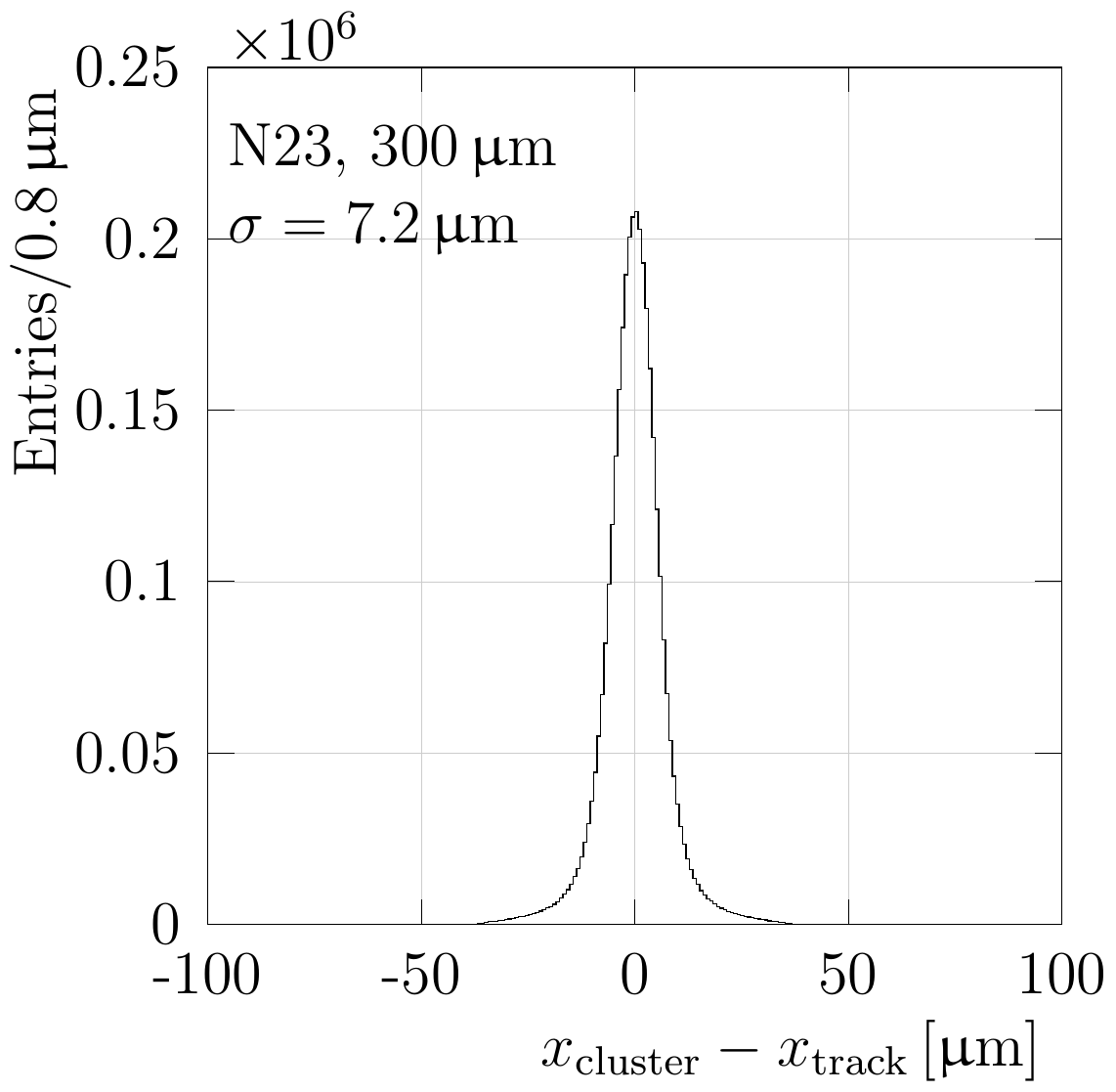}
  \includegraphics[width=0.48\textwidth
%	 ,natheight=400, natwidth =400
  ]{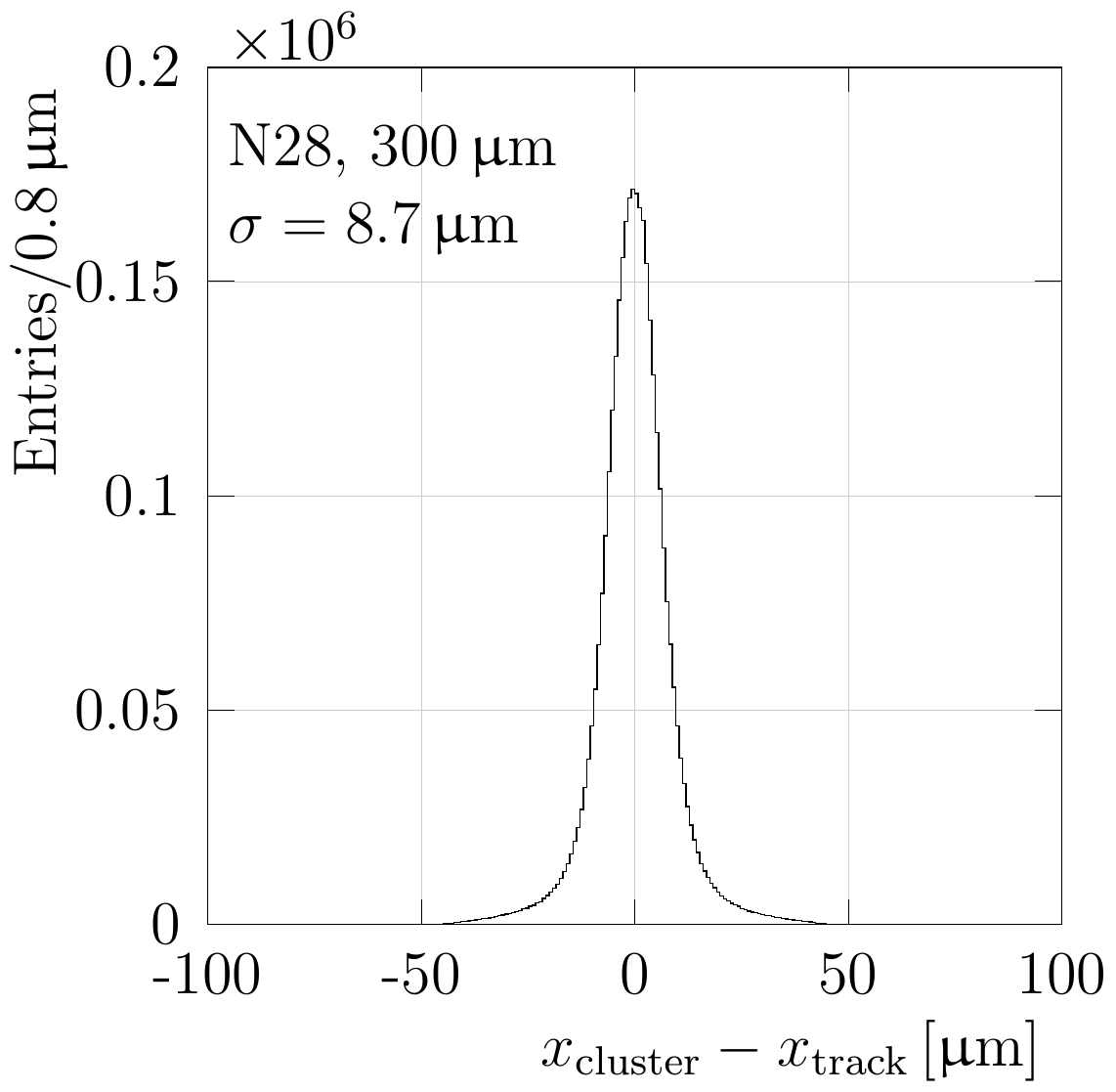}
  
  \vspace{-2mm}
  \caption{
  Distributions of $x$ residuals for the clusters of each plane.
  The residual is defined as the difference between the cluster position and the intercept of the associated track.
  }
  \label{fig:unbiased-residuals}
\end{figure}

\begin{figure}[tb]
  \centering
  \includegraphics[width=0.60\textwidth
%	 ,natheight=400, natwidth =400
  ]{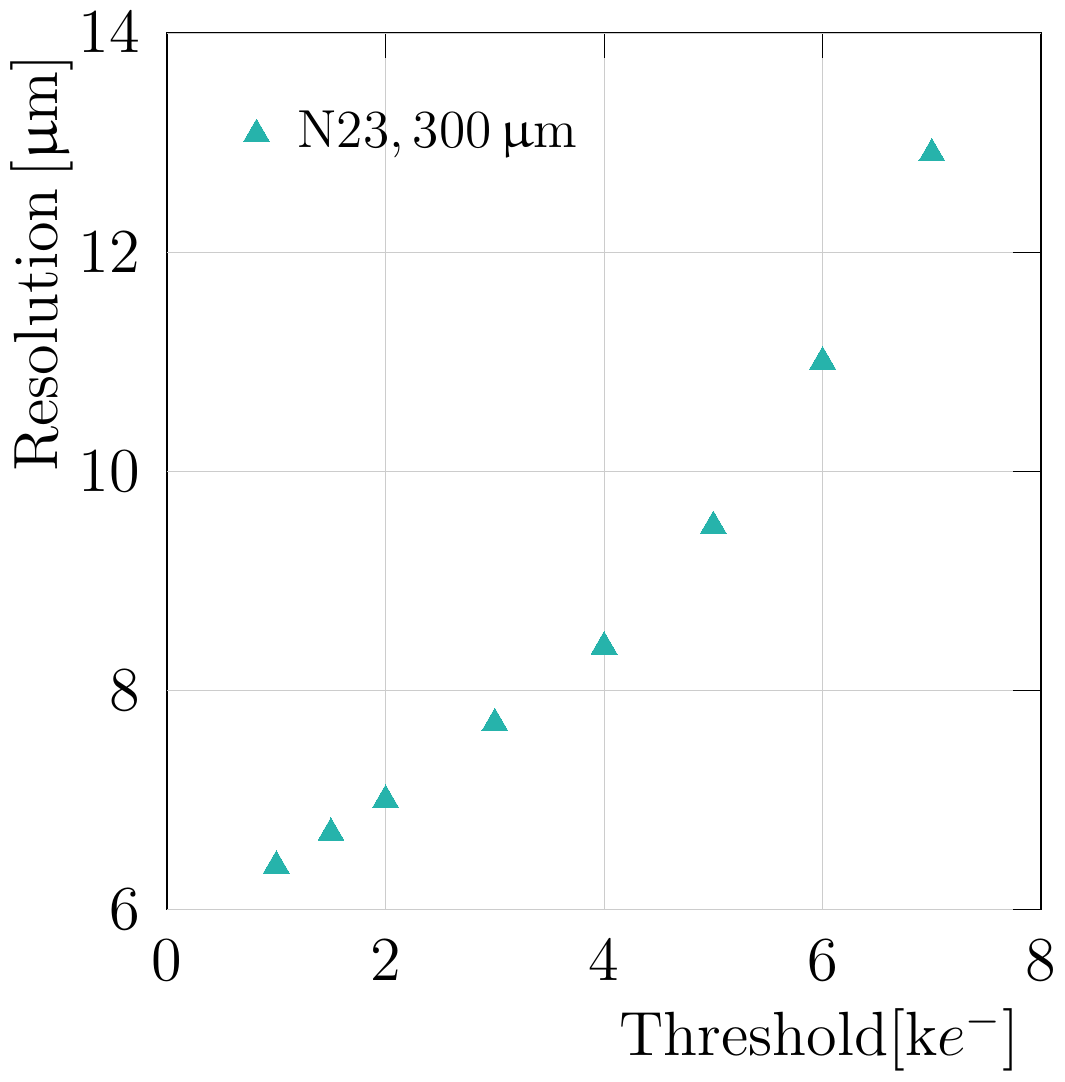}
  \vspace{-2mm}
  \caption{
  Resolution for the central $300 \mum$ plane as a function of threshold.
  }
  \label{fig:residuals-vs-thresh}
\end{figure}

The residual is defined as the difference between the position of a cluster, and the extrapolated position of the track to the given plane. 
The residual is said to be {\it unbiased} if the cluster on the plane of interest is excluded from the track fit. 
The unbiased residuals are determined in the ASIC coordinate system where the $x$ and $y$ axes correspond to the directions of increasing column and row numbers, respectively. 
The resulting distributions are shown in \cref{fig:unbiased-residuals}.
The spatial resolution of each plane is defined as the RMS of the unbiased residuals.
Clusters outside of a central interval containing 99\% of the distribution are discarded before calculating the RMS, which is then referred to as the truncated RMS. 
The $x$ residuals for the nominal data-taking conditions are shown in \cref{fig:unbiased-residuals}.
The truncated RMS is found, with negligible uncertainty, to be $33.2\mum, 16.6\mum, 7.2\mum$ and $8.7\mum$ for N30, N29, N23 and N28, respectively. 
The residual distribution is given by the convolution of the intrinsic resolution of the detector and the resolution of the track projection.
The latter is the dominant contribution to the residual on the first plane due to the long extrapolation distance, and is estimated to be around $30\mum$ from the track fit. 
The majority of clusters consists of a single hit for the $100\mum$ planes placed perpendicular to the beam, which results in a worse resolution with respect to the angled planes.
This can be seen from the characteristic top-hat distribution of N29 shown in the top right of \cref{fig:unbiased-residuals}.
The intrinsic resolution of the planes at their operating tilt is estimated from simulation, assuming that the resolution is equal in each direction and identical for planes with the same thickness and tilt.
The resolutions are found to be $(15.5 \pm  0.5) \mum$ for N30 and N29 and  $(4.5\pm 0.3) \mum$ for N23 and N28, in agreement with the values found for tilted $300 \mum$ sensors bonded to Timepix3~\cite{Akiba:2019faz}.
The resolution  degrades significantly with increasing operating threshold, as the measurement small charges becomes progressively lost, deteriorating the interpixel positon determination. This effect is shown in \cref{fig:residuals-vs-thresh}. Conversely, the resolution is found to be largely independent of the applied bias voltage. 

The single-plane efficiency is measured for each plane by reconstructing tracks from the other three planes and by searching for a cluster within $150\mum$ and $100\ns$ in space and time, respectively.
The efficiencies are found to be  $(92.0 \pm 5.0)\%$, $(99.4 \pm 0.2)\%$, $(99.1 \pm 0.4)\%$ and $(98.2 \pm 0.3)\%$ for planes N30, N29, N23 and N28, respectively.
The uncertainties are assigned using run-to-run variations throughout the data taking period.
The smaller efficiency and larger variation for plane N30 is due to a large number (around 10\%) of 
malfunctioning columns.

%% file: 06_temporalPerformance.tex
In this section, the temporal performance of each of the four Timepix4 planes is assessed. 
The \ttt (TtT) is defined as the difference between the timestamp of the earliest hit in a cluster and the reference time. 
The \ttt is analogous to the residuals for the spatial measurements, which yields the main figure-of-merit used in this section, the temporal resolution, defined as the RMS of the \ttt distribution. 
The timestamps are corrected for timewalk and per-pixel time offsets.
After applying these corrections, the resolution is studied as a function of bias and threshold.

\subsection*{Timewalk correction}

It is important to correct for timewalk for low-amplitude signals, such as from the 100\mum sensors or for hits that share charge with other pixels in the same cluster in the 300\mum planes.
The timewalk correction is performed based on the ToT of each hit, instead of the measured charge, since an accurate charge calibration procedure has not been developed for Timepix4 yet.

\begin{figure}  
  \centering
  \includegraphics[width=0.65\textwidth]{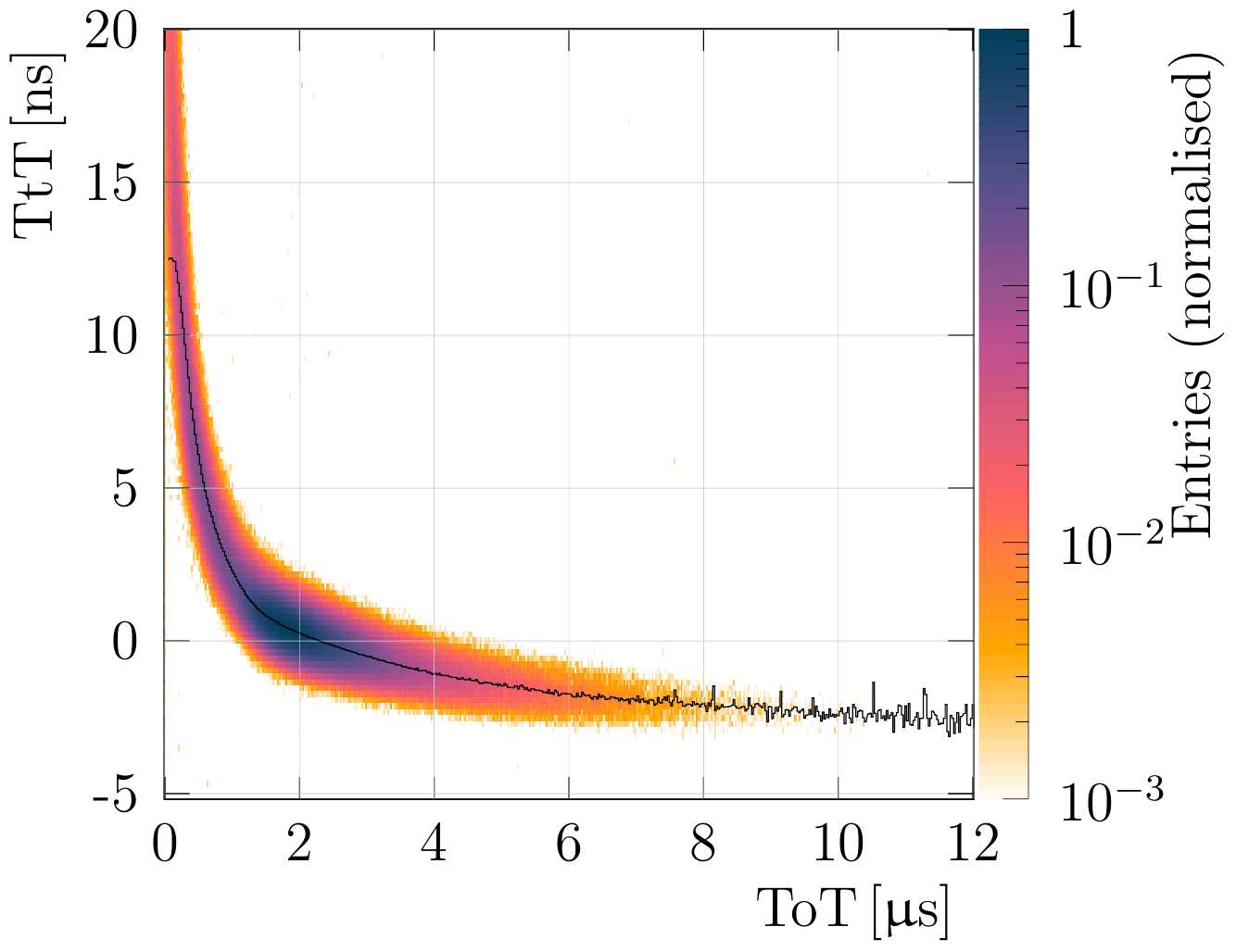}
  \includegraphics[width=0.65\textwidth]{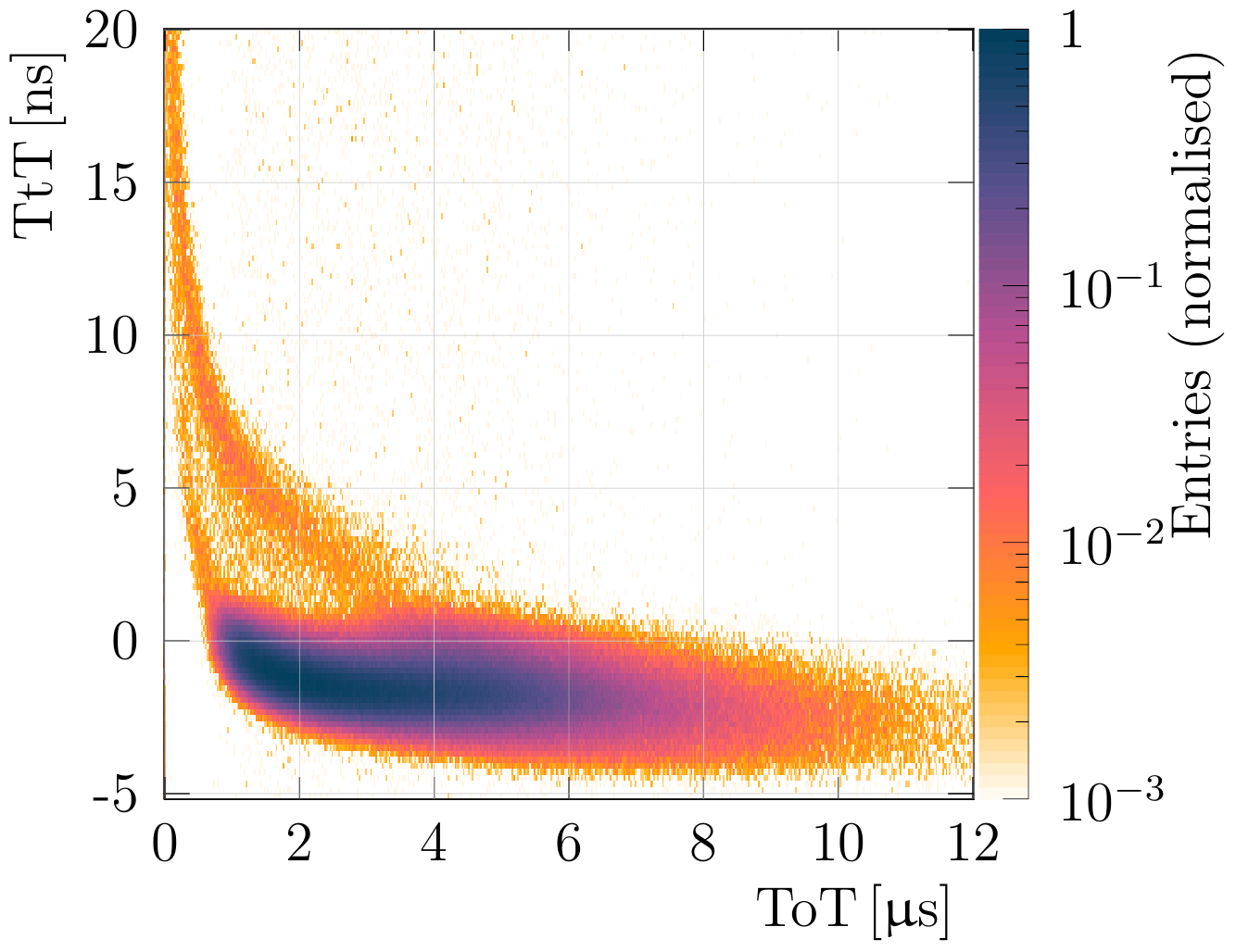}

  \caption{Top (bottom): Typical timewalk distribution for a $100\mum$ ($300\mum$) plane biased at 50 V (130 V).
  Since the $300\mum$ plane is tilted, the typical timewalk distribution shows multiple bands. The black line indicates the average of the distribution. 
  %(N30 and N23 run 122)
  }
  \label{fig:time:timewalk}
\end{figure}

Two different timewalk correction methods are employed, depending on the angle of the sensor with respect to the beam, as described in~\cite{Heijhoff:2020mlk}.
For the perpendicular (100\mum) sensors, the timewalk correction is performed exclusively using the ToT of hits.
A lookup table that contains the average TtT for each value of ToT is created per plane.
An example timewalk distribution for N29 ($100\mum$) is shown in \cref{fig:time:timewalk}~(top), where the  line indicates the values in the lookup table.
For the tilted (300\mum) sensors, the correction needs to account for timewalk and drift times, since the charge carriers can be liberated at different distances to the pixel implants~\cite{Heijhoff:2020mlk}. 

The timewalk distribution for a tilted sensor is shown in \cref{fig:time:timewalk}~(bottom).
Multiple bands can be seen in the distribution, 
indicating the necessity of a correction that additionally accounts for the intrapixel track position at each plane.
This method is described in detail in ref.~\cite{Heijhoff:2020mlk}.
Since this correction depends on drift velocity and threshold,
the lookup table is determined for each set of operational settings.

\subsection*{Per-pixel corrections}
A correction is required to account for per-pixel time offsets
due to differences in VCO start time and VCO frequency variations. 
The average TtT is determined for each pixel 
to account for these differences.
Corrections for differences of the TDC bin sizes are not implemented due to the limited size of the data samples.

\begin{figure}[tb]
  \centering
  {\includegraphics[width=0.65\textwidth]{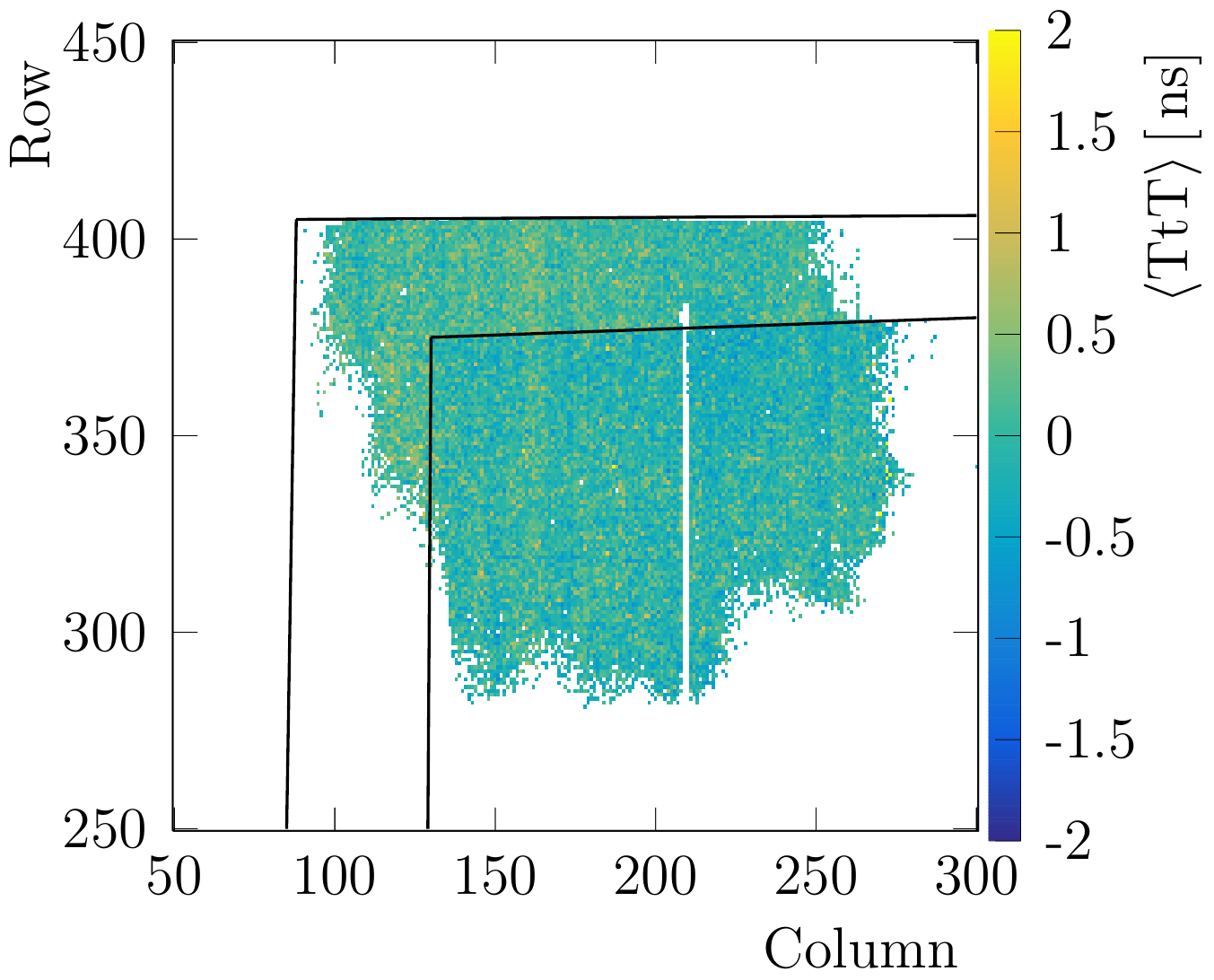}}
  \caption{Measured average TtT of individual pixels of N29. The lines indicate the regions covered by each scintillator. The absence of the vertical line of data is due to partially malfunctioning columns. 
  }
  \label{fig:time:delayMatrix}
\end{figure}
\Cref{fig:time:delayMatrix} shows the average TtT of the pixels of N29 ($100\mum$), where the lines indicate the regions covered by the two upstream scintillators.
The timestamps are corrected for timewalk before the average is determined for each pixel. 
The distribution of the average TtT of these pixels shows a large variation with an RMS of $315\ps$.
This effect is corrected by subtracting the average TtT of the pixel from the timestamp.

\subsection*{Time resolution}
\begin{figure}[tb]
  \centering
  {\includegraphics[width=0.48\textwidth]{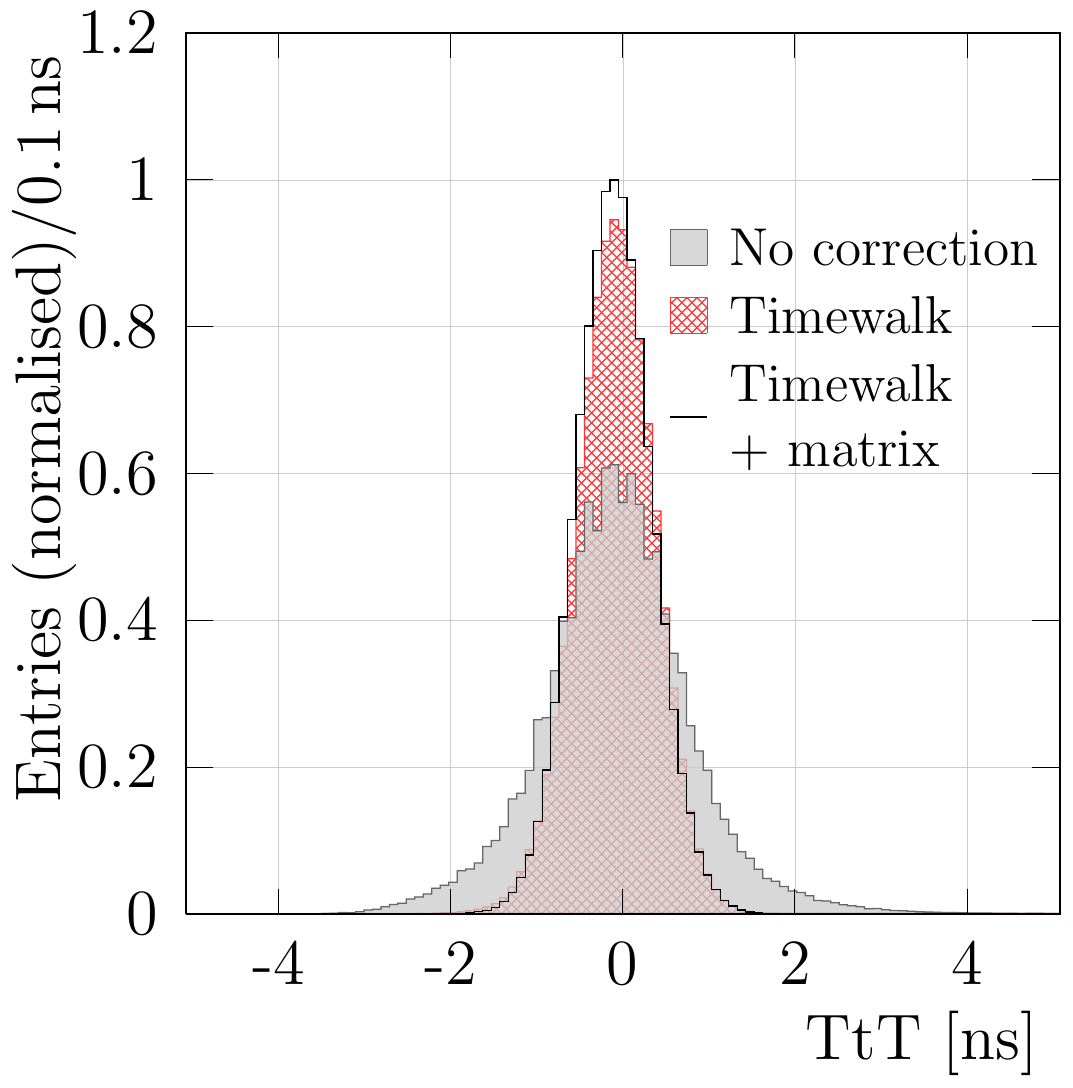}}
  \caption{TtT distribution of all pixels of N29 ($100\mum$) biased at 50 V and a threshold of \mbox{$1000 \, \en$}. 
  The filled histogram indicates the uncorrected TtT distribution, 
  the hashed represents this distribution after the timewalk correction, and the solid line displays this distribution after both a timewalk and per-pixel time offset correction.
  }
  \label{fig:time:temporalTimeRes}
\end{figure}

The four planes of the telescope are characterised  as a function of the bias voltages and threshold. 
The temporal resolution is determined after both  timewalk and per-pixel time offset corrections have been applied. 
\Cref{fig:time:temporalTimeRes} shows the TtT distribution, before any correction (filled histogram), after the timewalk correction (hashed), and after both timewalk and per-pixel delay corrections (solid line).
The time resolution is improved from $(783\pm24)\ps$ to $(439\pm10)\ps$, implying that a total contribution of $648\ps$ has been removed.

\begin{figure}[tb]
  \centering
  \includegraphics[width=0.48\textwidth]{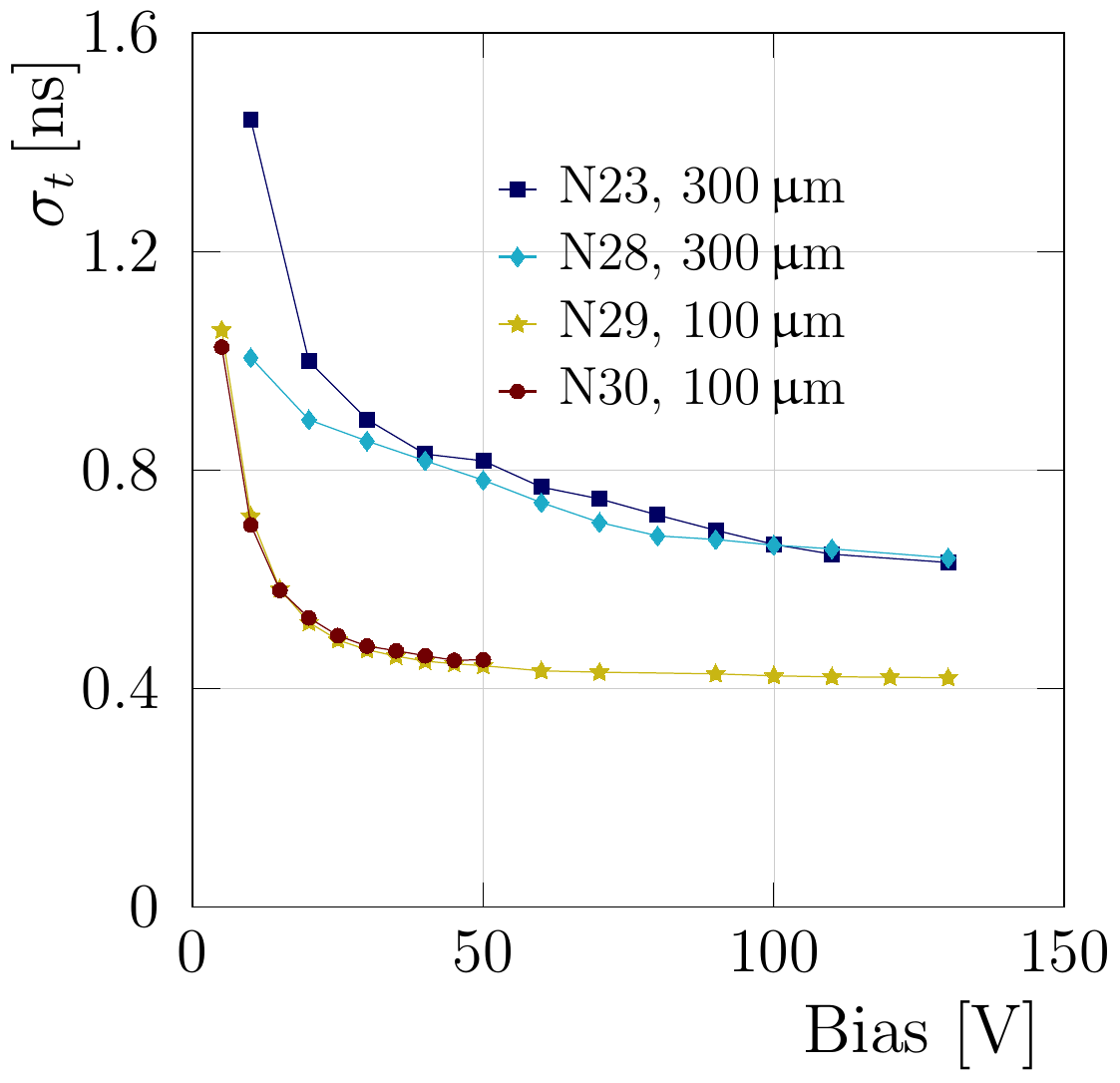} \includegraphics[width=0.48\textwidth]{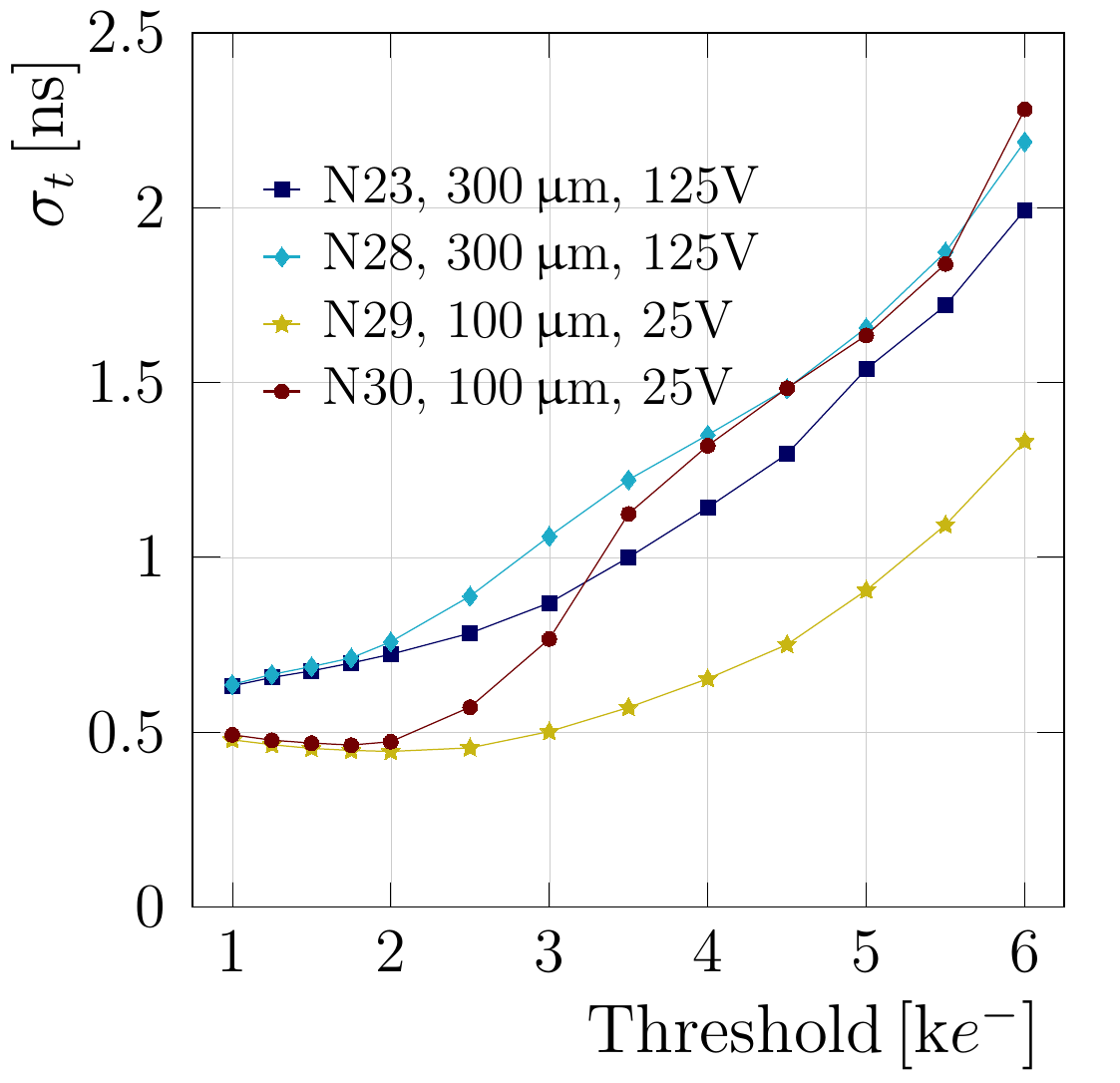}
  \caption{Left (right): time resolution of the four telescope planes as a function of the bias voltage (threshold). The four planes are indicated by the different markers. }
  \label{fig:time:temporalScans}
\end{figure}

The resolution changes as a function of operational settings such as bias voltage and threshold. 
Scans over these parameters are shown in \cref{fig:time:temporalScans}, where the left and right figures show bias and threshold scan, respectively. 
For all planes the time resolution shows improvement for higher bias voltages.
The two tilted $300\mum$ sensors have a resolution that is significantly worse than that of the $100\mum$ sensors.
The main cause is the more complex timewalk correction in addition to higher variations in the Ramo-Shockley weighting field, in comparison to the $100\mum$ sensors.

As shown in \cref{fig:time:temporalScans} (right) 
the time resolution slightly improves as a function of threshold for the two $100\mum$ sensors, reaching the best resolution around the value of \mbox{$2000\, \en$}. This effect is counterintuitive but well described in ref.~\cite{tpx4_jinst}.
The two tilted $300\mum$ sensors do not show a local minimum. This is probably due to the larger variations in the time corrections. Plane N30 achieves its best resolution at 50~V, and the other planes at 130~V, all at a threshold of \mbox{$1000 \, \en$}. Their time resolutions are $(452\pm10)\ps$, $(420\pm10)\ps$, $(639\pm10)\ps$, and $(631\pm10)\ps$ for N30, N29, N38, N23, respectively. The uncertainty is estimated from run-to-run variations. The resolution of the detectors saturates around 50~V for the 100\mum sensors and around 130~V for the 300\mum sensors, and an improvement by increasing the bias voltage further is not expected. 

The track time is determined by taking the uncertainty-weighted average of the individual measurements.
To achieve the best track time resolution the planes should be biased at the highest operational high voltages. 
The resolution is determined in a configuration where the two thin planes are biased at 50~V and the other planes at 130~V, while the 
threshold is \mbox{$1000\, \en$} for all planes. 
The achieved track resolution is $(340\pm5)\ps$. 
The result of the combination of single plane measurements to a track is worse than what is expected from the na\"ive  calculation using the separate resolutions.
This is due to correlations between the time measurements, which can lead to a significantly worse resolution~\cite{Heijhoff:2020mlk}, and drift in the synchronisation between the planes.

%% file: 07_conclusion.tex
A system composed of four Timepix4 detectors is used to reconstruct high energy hadrons from the CERN SPS H8 beam line. 
The overall spatial resolution is assessed for each of the detector planes by projecting the reconstructed tracks using the other three planes. 
The resolutions in the default configuration are estimated to be $(15.5\pm 0.5)\mum$ and $(4.5\pm0.3)\mum$ for $100\mum$ and $300\mum$ thick sensors, respectively, after subtracting the expected contribution from the track extrapolation.

The timestamps from the detector are corrected for timewalk and per-pixel time offsets, finally yielding individual time resolutions of $(452\pm10)\ps$, $(420\pm10)\ps$, $(639\pm10)\ps$, and $(631\pm10)\ps$ for N30 (perpendicular 100\mum), N29 (perpendicular 100\mum), N28 (tilted 300\mum), and N23 (tilted 300\mum), respectively, when compared to the measurements from the reference scintillators.
These resolutions have been achieved at a threshold of \mbox{$1000\, \en$} and 50~V bias for N30, and 130~V for the other planes.
These measurements can be combined to a track time resolution of $(340\pm5)\ps$.

%% file: 08_acknowledgements.tex
We would like to express our gratitude to our colleagues in the CERN accelerator departments for the excellent performance of the beam in the SPS North Area. 
We gratefully acknowledge the support of the CERN Strategic R\&D Programme on Technologies for Future Experiments\footnote{https://ep-rnd.web.cern.ch/} and the computing resources provided by CERN. 
We also gratefully acknowledge the support from the following national agencies: the Netherlands Organisation for Scientific Research (NWO); The Royal Society and the Science and Technology Facilities Council (U.K., grant no. ST/V003151/1 and ST/S000933/1); 
the European Research Council (grant no. 852642);
the Wolfgang Gentner Programme of the German Federal Ministry of Education and Research (grant no. 13E18CHA);
the German Federal Ministry of Education and Research (BMBF, grant no. 05H21PECL1) within ErUM-FSP T04;
and the European Union’s Horizon 2020 Research and Innovation programme (GA no. 101004761).

%% file: main.bbl
\providecommand{\href}[2]{#2}\begingroup\raggedright\begin{thebibliography}{10}

\bibitem{LHCbVELOgroup:2022}
K.~Akiba et~al., \emph{{Considerations for the VELO detector at the LHCb
  Upgrade II}}, {\emph{LHCb Public Note: CERN-LHCb-PUB-2022-001} (2022) }.

\bibitem{Akiba:2019faz}
K.~Akiba et~al., \emph{{LHCb VELO Timepix3 Telescope}},
  \href{https://doi.org/10.1088/1748-0221/14/05/P05026}{\emph{JINST} {\bfseries
  14} (2019) P05026} [\href{https://arxiv.org/abs/1902.09755}{{\ttfamily
  1902.09755}}].

\bibitem{Heijhoff:2020mlk}
K.~Heijhoff et~al., \emph{{Timing performance of the LHCb VELO Timepix3
  Telescope}},
  \href{https://doi.org/10.1088/1748-0221/15/09/p09035}{\emph{JINST} {\bfseries
  15} (2020) P09035} [\href{https://arxiv.org/abs/2008.04801}{{\ttfamily
  2008.04801}}].

\bibitem{Heijhoff:2021rtu}
K.~Heijhoff et~al., \emph{{Timing measurements with a 3D silicon sensor on
  Timepix3 in a $180\gevc$ hadron beam}},
  \href{https://doi.org/10.1088/1748-0221/16/08/P08009}{\emph{JINST} {\bfseries
  16} (2021) P08009} [\href{https://arxiv.org/abs/2105.11800}{{\ttfamily
  2105.11800}}].

\bibitem{DallOcco:2021tjb}
E.~Dall'Occo et~al., \emph{{Temporal characterisation of silicon sensors on
  Timepix3 ASICs}},
  \href{https://doi.org/10.1088/1748-0221/16/07/P07035}{\emph{JINST} {\bfseries
  16} (2021) P07035} [\href{https://arxiv.org/abs/2102.06088}{{\ttfamily
  2102.06088}}].

\bibitem{tpx4_jinst}
X.~Llopart et~al., \emph{Timepix4, a large area pixel detector readout chip
  which can be tiled on 4 sides providing sub-200 ps timestamp binning},
  \href{https://doi.org/10.1088/1748-0221/17/01/c01044}{\emph{JINST} {\bfseries
  17} (2022) C01044}.

\bibitem{llopart2022timepix4}
X.~Llopart, J.~Alozy, R.~Ballabriga, M.~Campbell, R.~Casanova, V.~Gromov
  et~al., \emph{Timepix4, a large area pixel detector readout chip which can be
  tiled on 4 sides providing sub-200 ps timestamp binning}, {\emph{Journal of
  Instrumentation} {\bfseries 17} (2022) C01044}.

\bibitem{sps-h8}
{CERN Engineering Department}, ``{The H8 Secondary Beam Line of EHN1/SPS}.''

\bibitem{emma}
E.~Buchanan et~al., \emph{Spatial resolution and efficiency of prototype
  sensors for the {LHCb} {VELO} upgrade},
  \href{https://doi.org/10.1088/1748-0221/17/06/p06038}{\emph{JINST} {\bfseries
  17} (2022) P06038} [\href{https://arxiv.org/abs/2201.12130v2}{{\ttfamily
  2201.12130v2}}].

\bibitem{Clemencic_2010}
M.~Clemencic et~al., \emph{Recent developments in the {LHCb} software framework
  gaudi}, \href{https://doi.org/10.1088/1742-6596/219/4/042006}{\emph{Journal
  of Physics: Conference Series} {\bfseries 219} (2010) 042006}.

\bibitem{Gaspar:2001fbw}
C.~Gaspar, M.~D\"onszelmann and P.~Charpentier, \emph{{DIM, a portable, light
  weight package for information publishing, data transfer and inter-process
  communication}},
  \href{https://doi.org/10.1016/S0010-4655(01)00260-0}{\emph{Comput. Phys.
  Commun.} {\bfseries 140} (2001) 102}.

\bibitem{Blobel:2006yh}
V.~Blobel, \emph{{Software alignment for tracking detectors}},
  \href{https://doi.org/10.1016/j.nima.2006.05.157}{\emph{Nucl. Instrum. Meth.
  A} {\bfseries 566} (2006) 5}.

\end{thebibliography}\endgroup
